\documentclass[prd,preprint,amsmath,amssymb,nofootinbib,eqsecnum]{revtex4-1}
\usepackage{latexsym,graphics,color,epsfig,hyperref,mathrsfs}

\newcommand{\ie}{{i.e.}}
\newcommand{\cf}{{cf.}}

\newcommand{\viz}{{viz.}}
\newcommand{\wrt}{with respect to}
\newcommand{\lhs}{left-hand side}

\newcommand{\rhs}{right-hand side}

\newcommand{\Naively}{Na\"{\i}vely}

\newcommand{\be}{\begin{equation}}
\newcommand{\ee}{\end{equation}}
\newcommand{\bse}{\begin{subequations}}
\newcommand{\ese}{\end{subequations}}

\newcommand{\bcf}{\begin{center}\begin{figure}}
\newcommand{\ecf}{\end{figure}\end{center}}

\newcommand{\bct}{\begin{center}\begin{table}}
\newcommand{\ect}{\end{table}\end{center}}

\newcommand{\eq}[1]{(\ref{eq:#1})}

\newcommand{\Eqn}[1]{Equation~(\ref{eq:#1})}

\newcommand{\eqs}[2]{(\ref{eq:#1}) and~(\ref{eq:#2})}
\newcommand{\eqss}[3]{(\ref{eq:#1}), (\ref{eq:#2}) and~(\ref{eq:#3})}

\newcommand{\sect}[1]{section~\ref{sec:#1}}
\newcommand{\Sect}[1]{Section~\ref{sec:#1}}

\newcommand{\app}[1]{appendix~\ref{app:#1}}

\newcommand{\D}{d}
\newcommand{\Int}[1]{\int \!\! d^{\D} \! #1 \,}
\newcommand{\CurvedInt}[1]{\int \!\! d^{\D} \! #1 \sqrt{g} \,}
\newcommand{\MomInt}[1]{\int \!\! \frac{d^{\D} \! #1}{(2\pi)^{\D}} \,}

\newcommand{\LimitInt}[3]{\int_{#2}^{#3} \! d #1 \,}

\newcommand{\DD}[1]{\delta^{(#1)}}

\newcommand{\measure}[1]{\bigl[ d #1 \bigr]}

\newcommand{\der}[2]{\frac{d #1}{d #2}}

\newcommand{\fder}[2]{\frac{\delta #1}{\delta #2}}
\newcommand{\fderhat}[2]{\frac{\hat{\delta} #1}{\delta #2}}

\newcommand{\dfderhat}[3]{\frac{\hat{\delta}^2 #1}{\delta #2 \, \delta #3}}

\newcommand{\Or}{\mathrm{O}}
\newcommand{\order}[1]{\Or \bigl( #1 \bigr)}

\newcommand{\hf}{\frac{1}{2}}
\newcommand{\smallhf}{{\textstyle \frac{1}{2}}}

\newcommand{\eval}[1]{
	\bigl\langle #1 \bigr\rangle
}

\newcommand{\one}{1\!\mathrm{l}}

\newcommand{\Tr}{\mathrm{Tr}\,}

\newcommand{\Ricci}{R}

\newcommand{\SCT}{\mathcal{K}}
\newcommand{\Dil}{\mathcal{D}}

\newcommand{\dil}[1]{D^{(#1)}}
\newcommand{\sct}[2]{{K^{(#1)}}_{#2}}

\newcommand{\qsource}{J}

\newcommand{\emtLegendre}{\Theta}

\newcommand{\scaling}[1]{\delta_{#1}}

\newcommand{\fp}[1]{#1}

\newcommand{\weyl}{\omega}
\newcommand{\Weyl}{\mathrm{Weyl}}

\newcommand{\WeylLaplacian}{\Delta_\mathrm{W}}

\newcommand{\Lie}[1]{\mathcal{L}_{#1}}

\newcommand{\anomaly}{\mathcal{A}}

\newcommand{\pf}{Z}
\newcommand{\field}{\phi}
\newcommand{\dfield}{\varphi}
\newcommand{\efield}{\Phi}

\newcommand{\cutoff}{K}
\newcommand{\ep}{\mathcal{G}}
\newcommand{\G}{G}
\newcommand{\RIR}{\mathscr{R}}

\newcommand{\Stot}{S}

\newcommand{\Gint}{\Gamma}
\newcommand{\SGauss}{\Stot^{\mathrm{Gauss}}}

\newcommand{\R}{\varrho}
\newcommand{\Rinv}{\sigma}

\newcommand{\Schwinger}{W}

\newcommand{\EMT}{T}

\newcommand{\linear}[1]{\mathcal{C}_{#1}}
\newcommand{\quadratic}[2]{\mathcal{Q}_{#1,#2}}

\newcommand{\classical}[1]{{#1}^{\mathrm{cl}}}

\newcommand{\pskernel}[2]{{#1} \cdot \bigl(G \otimes \one + \one \otimes G\bigr) \cdot {#2}}
\newcommand{\kernel}[2]{{#1} \cdot \bigl(G \times \one + \one \times G\bigr) \cdot {#2}}

\newcommand{\hepth}[1]{hep-th/#1}
\newcommand{\hepph}[1]{hep-ph/#1}

\begin{document}

\title{The Conformal Anomaly and a new Exact RG}

\author{Oliver J.~Rosten}
\email{oliver.rosten@gmail.com}
\affiliation{Unaffiliated}

\begin{abstract}
	For scalar field theory, a new generalization of the Exact RG to curved space is proposed,
	in which the conformal anomaly is explicitly present.
	Vacuum terms require regularization beyond that present in the 
	canonical formulation of the Exact RG, which can be 
	accomplished by adding certain free fields, each at a 
	non-critical fixed-point.
	Taking the Legendre transform, the sole 
	effect of the regulator fields is to remove a divergent vacuum 
	term and they do not explicitly appear in the effective average 
	action. As an illustration, both the integrated conformal anomaly and Polyakov action
	are recovered for the Gaussian theory in $\D=2$.
	\vspace{8ex}

	\begin{center}
		\large \em{In memory of David Bailin}
	\end{center}
\end{abstract}

\maketitle
\tableofcontents

\section{Introduction}

Conformal Field Theory (CFT) and Wilsonian renormalization are central to the modern understanding of
quantum field theory. However, the mathematical understanding of CFTs has reached a considerably higher level of maturity, particularly in $\D=2$. Indeed, by comparison, the mathematical formulation of Wilson's ideas---the Exact Renormalization Group (ERG)---remains something of a niche area. This is striking when one considers that Wilson's picture of theory space filled with RG trajectories and enriched by fixed-points is part of the conceptual fabric of quantum field theory.

The original motivation for Wilson's work~\cite{Wilson} was to understand systems with local interactions in the difficult regime where there are very many degrees of freedom within a correlation length. A crucial piece of insight came from Kadanoff's blocking procedure~\cite{Kadanoff}: by averaging in some way over local patches, one arrives at a coarse-grained description appropriate to a larger distance scale than the original microscopic picture. Iterating this process, one can ultimately hope to gain an understanding at distances of the order of the correlation length, thereby solving the original problem.

Combining the Kadanoff procedure with a subsequent rescaling of the system constitutes the renormalization procedure envisaged by Wilson. Considering fixed-points of this process and linearization in their vicinity gives a compelling qualitative understanding of universality, asymptotic freedom, asymptotic safety and triviality (see~\cite{Fundamentals} for a detailed description and further references). The ERG provides a framework which offers a quantitative approach to such physics though, as with any attempt to understand non-perturbative quantum field theory, extracting reliable predictions in a controlled fashion is very challenging.

Focussing on theories of a single scalar field, $\dfield$, let us introduce the RG time, $t$, upon which the Wilsonian effective action is taken to depend, \viz\ $\Stot_t[\dfield]$. The direction of increasing $t$ corresponds to the evolution from the ultraviolet (UV) to the infrared (IR). The ERG equation describes how the Wilsonian effective action behaves under an infinitesimal change of scale, for which there are several contributions. 
 Denoting the scaling dimension of the field by $\scaling{\dfield}$ (which may depend on $t$) and encoding the details of the blocking transformation in $\Psi$, the general form is:
\be
	\biggl(
		\partial_t + 
		\dil{\scaling{\dfield}} \dfield \cdot \fder{}{\dfield}
	\biggr)
	e^{-\Stot_t[\dfield]}
	+	
	\fder{}{\dfield}
	\cdot 
	\Bigl(
		\Psi
		e^{-\Stot_t[\dfield]}
	\Bigr)
	= 0
.
\label{eq:schematic}
\ee

Several comments are in order. The second term generates dilatations and may be written as follows:
\be
	\dil{\scaling{\dfield}} \dfield \cdot \fder{}{\dfield}
	\equiv
	\Int{x} \bigl(x^\mu \partial_\mu + \scaling{\dfield}\bigr)
	\dfield(x)
	\fder{}{\dfield(x)}
,
\label{eq:dil-functional}
\ee
where it is apparent that, in this context, the dot is a shorthand for integration over all space.
By considering the effect on typical contributions to the action, such as the kinetic term
$\partial_\mu \dfield \cdot \partial^\mu \dfield$, or the mass term $\dfield \cdot \dfield$, a piece of intuition follows, valid classically: this dilatation generator counts the scaling dimension, with $-\D$ coming from the integral over all space, each space-time derivative being counted with a weight of unity, and each instance of the field being weighted by the scaling dimension, $\scaling{\dfield}$. However, it should be borne in mind that, for interacting theories, there are quantum corrections arising from the blocking term. As discussed below, this effectively means that while still present in the ERG equation, \eq{dil-functional} ceases to be the appropriate representation of the dilatation generator.

Returning to~\eq{schematic}, the precise relationship between $\Psi$ and the blocking kernel which implements averaging over local patches is straightforward and may be found in~\cite{mgierg1} (see also~\cite{Fundamentals}). Note from the form of the blocking contribution that it may be interpreted as a field redefinition, which has been explored in depth notably in~\cite{TRM+JL}. There is considerable freedom in the blocking procedure and hence a commensurate freedom in the form of the ERG equation. However, for the purposes of this paper it is sufficient to restrict to the following
\be
	\Psi = A \cdot \dfield + B \cdot \fder{}{\dfield}
,
\ee
where $A = A(x,y)$ is independent of the field (ditto $B$) and, consistently with above, the dot indicates an integral over the shared coordinate:
\be
	\bigl(A \cdot \dfield \bigr)(x)
	=
	\Int{y} A(x,y)  \dfield(y) ,
	\qquad
	\bigl(
		A \cdot A^{-1}
	\bigr)(x,y)
	=
	\DD{\D}(x-y)
.
\label{eq:dot}
\ee
A feature which all legitimate ERG equations share is quasi-locality: ingredients such as $A$ and $B$ must exhibit a derivative expansion, as a consequence of the requirement that the blocking procedure is over local patches. Specific choices of $A$ and $B$ generate ERG equations structurally similar to those of Wilson or Polchinski~\cite{Pol}; for the purposes of this paper, the canonical ERG equation of~\cite{Ball} will be employed.

One of the problems from which the ERG suffers is a paucity of exact results (which is perhaps an unfortunate irony); a second, which is not entirely unrelated, is that it is not always immediately obvious how the ERG relates to more standard approaches. Nevertheless, it was not long after the discovery of the ERG~\cite{Wilson} that Sch\"afer recognized its intimate relationship with CFT~\cite{Schafer-Conformal}. It is curious that, until recently, this has not been much explored.

An attempt has been made to partially rectify this deficiency in the sequence of papers~\cite{Representations,WEMT,CFPE,WWID} (see also~\cite{Delamotte-Conformal,Sonoda:ERG/CFT}). Two main results have come out of this. First is that, for CFTs, there is a precise sense in which the ERG furnishes a representation of the conformal algebra. Secondly, the trace of the energy-momentum tensor, $\EMT(x)$, plays a fundamental role in the structure of the ERG. In particular, $\EMT$ is none other than the exactly marginal, redundant field which has long been known to exist at every critical fixed-point~\cite{WegnerInv,RGN}.

The focus will now shift to CFTs and hence to fixed-point solutions of the ERG, defined by $\partial_t \Stot[\dfield] = 0$. Henceforth, unless stated otherwise, we implicitly operate in this domain and so drop the $t$ from $\Stot$. 
The structure of $\EMT$  within the (canonical) ERG has some notable features. Introducing a UV cutoff function, $\cutoff$, it takes the form
\be
	T e^{-\Stot}
	= 
	\scaling{\dfield}
	\fder{}{\dfield} \cdot \cutoff \times
	\biggl(
		\cutoff^{-1} \cdot \dfield
		+ \R \cdot \fder{}{\dfield}
	\biggr)
	e^{-\Stot}
,
\label{eq:trace_EMT}
\ee
where an expression for $\R$ will be given later and the multiplication symbol is used, where appropriate, to emphasise the lack of a dot. As such, \eq{trace_EMT} corresponds to the unintegrated $T(x)$. 
The first thing to observe about the expression for $\EMT$ is that it contains a divergent contribution arising from the first functional derivative striking the adjacent instance of $\dfield$.%
\footnote{Since vacuum terms have been largely ignored in the past, this issue has been previously glossed over.}
In principle, the divergence may be cancelled by a term arising from the two functional derivatives acting in concert to reduce a two-point contribution to $\Stot$ to a vacuum term; indeed, an example of this will be presented later. Moreover, upon taking a Legendre transform, it will be apparent that this cancellation is in fact guaranteed. However, at this stage, to be rigorous we should exploit the factorization of $\EMT$ apparent in~\eq{trace_EMT} to define
\be
	\EMT(x, y) e^{-\Stot}
	= 
	\scaling{\dfield}
	\biggl(
		\fder{}{\dfield} \cdot \cutoff
	\biggr)(x)
	\biggl(
		\cutoff^{-1} \cdot \dfield
		+ \R \cdot \fder{}{\dfield}
	\biggr)(y)
	\;
	e^{-\Stot}
,
\label{eq:T(x,y)}
\ee
with
\be
	\EMT(x) = \lim_{y \rightarrow x} \EMT(x,y)
,
\label{eq:T(x)}
\ee
should the limit exist.

The second thing to observe is that $\EMT$ is a total functional derivative. This implies that $\EMT$ may be generated by an infinitesimal field redefinition, confirming the claim above that it is `redundant',  
and therefore its expectation value always vanishes. In flat space, where the expression for $\EMT$ was derived, this is unsurprising. However, it begs the question as to what happens on a curved manifold due to the conformal anomaly. For an entertaining perspective on the latter, see~\cite{DuffTwentyYears}.

To answer this question suggests that some sort of generalization of the ERG to curved space is required. The simplest route, which will be primarily pursued in this paper, is to consider global Weyl transformations. This corresponds to a generalization of the standard ERG equation (or its Legendre transform), which encodes dilatation invariance. However, as recognized by Sch\"afer, the ERG equation has a partner which encodes invariance under special conformal transformations. In~\cite{WEMT,CFPE} it was noticed that the ERG equation and its partner can be combined into a single `conformal fixed-point equation'. 

It will be useful to make contact with this, particularly to aid understanding the emergence of full conformal invariance in the flat-space limit. To this end, let us review the statement of conformal invariance in flat space. Classically, a conformally invariant action, $\Stot[\dfield]$, satisfies%
\footnote{These equations are written in terms of $e^{-\Stot}$ since, when transitioning to the ERG, they remain conveniently linear.}
\be
	\dil{\scaling{\dfield}} \dfield \cdot \fder{}{\dfield} e^{-\Stot[\dfield]} = 0,
	\qquad
	\sct{\scaling{\dfield}}{\mu} \dfield \cdot \fder{}{\dfield} e^{-\Stot[\dfield]} = 0,
	\qquad
\label{eq:classical_conformal}
\ee
where we recognize the appearance of the conventional differential operators
\be
	\dil{\scaling{\dfield}} = x \cdot \partial + \scaling{\dfield},
	\qquad
	\sct{\scaling{\dfield}}{\mu} = 2x_\mu \bigl( x\cdot \partial + \scaling{\dfield} \bigr) - x^2 \partial_\mu
.
\ee
The first of these has been encountered already in~\eq{dil-functional}, which is of course no coincidence (note that the dot has been overloaded to additionally represent contraction of indices).

In quantum field theory, conformal invariance of the correlation functions is similarly realized. Specifically, coupling the fundamental field to a source, $\qsource$, the correlation functions may be encapsulated in the Schwinger functional, $\Schwinger[\qsource]$. For conformal field theories, this satisfies
\be
	\dil{\D-\scaling{\dfield}} \qsource \cdot \fder{}{\qsource} e^{\Schwinger[\qsource]} = 0,
	\qquad
	\sct{\D-\scaling{\dfield}}{\mu} \qsource \cdot \fder{}{\qsource} e^{\Schwinger[\qsource]} = 0.
\label{eq:Schwinger_conformal}
\ee

However, for the action, the story is more subtle in quantum field theory due to the need to regularize. To proceed, it helps to recognize the functional differential operators in~\eq{classical_conformal}, together with the operators corresponding to translations and rotations, as furnishing a representation of the conformal algebra---as discussed in great detail in~\cite{Representations}. Whereas above the ERG was introduced in a physically motivated way, explicitly utilizing blocking, here its fixed-point version can be thought of in a different, more abstract manner. Concretely, supplement the two generators in~\eq{classical_conformal} by additional terms such that:
\begin{enumerate}
	\item The additional terms are quasi-local, a corollary of which is that in the limit that the cutoff is removed, the generators reduce to their original (classical) form.

	\item The resulting generators still satisfy the conformal algebra.
	
	\item The equation for $\Stot$ is a non-linear eigenvalue equation with each quasi-local solution self-consistently determining a value of $\scaling{\dfield}$.
\end{enumerate} 
The upshot of this is that conformal invariance may be realized in a non-linear manner by actions which incorporate a cutoff function. It is worth dwelling on this. \Naively, we would expect the presence of a cutoff function to break conformal invariance. This would not necessarily be a disaster since we may 
nevertheless expect conformal invariance to be recovered as a property of correlation functions computed from such actions. However, the situation is better than this: the ERG furnishes representations of the conformal generators---concrete forms of which will be given in \sect{Defining}. As such, we may usefully speak of a symmetry being \emph{directly} realized as a property of the Wilsonian effective action, by which we understand the following.  Given a classical functional representation of a symmetry generator, this symmetry is realized as a property of the classical action if the action is annihilated by the generator. If this generator may be deformed in a quasi-local way such that the deformed generator annihilates the Wilsonian effective action, then we will say that the symmetry is realized directly as a property of the Wilsonian effective action.

Motivated by the way in which conformal invariance is realized as a property of the Wilsonian effective action, the scheme we propose for constructing a curved space version of the ERG is to supplement the classical statement of Weyl invariance by additional quasi-local terms. Given a function $\weyl(x)$, Weyl transformations correspond to $g_{\mu\nu} \rightarrow g_{\mu\nu}e^{2\weyl}$.
For a classical action, emphasised by `cl', local Weyl invariance implies that
\be
	\classical{\Stot}[e^{-\scaling{\dfield}} \dfield, e^{2\weyl} g_{\alpha \beta}]
	=
	\classical{\Stot}[\dfield, g_{\alpha \beta}]
.
\ee
Infinitesimally,
\be
	\Int{x}
	\weyl(x)
	\biggl(		
		\scaling{\dfield} \, \dfield \, \fder{}{\dfield}
		- 2 g_{\mu\nu} \fder{}{g_{\mu\nu}}
	\biggr)
	e^{-\classical{\Stot}[\dfield, g_{\alpha\beta}]}
	= 0
.
\label{eq:Weyl-cl}
\ee
Note that the overall sign has been flipped for convenience, to match with the conventional choice for the flat space ERG equation.

Our hope is to generalize this latter equation to one of the form
\[
	\Int{x}
	\weyl(x)
	\biggl(		
		\scaling{\dfield} \, \dfield \, \fder{}{\dfield}
		- 2 g_{\mu\nu} \fder{}{g_{\mu\nu}}
		+ \cdots
	\biggr)
	e^{-{\Stot}[\dfield, g_{\alpha\beta}]}
	= 0
,
\]
where the ellipsis are chosen such that, upon restricting to flat space, appropriate choices of $\omega$ yield either the ERG equation or its special conformal partner.

As we will come to understand in detail, the presence of a cutoff function prevents us from achieving this for arbitrary $\omega$. Consequently, for much of this paper we will focus on global Weyl transformations. Global Weyl invariance may be realized directly as a property of the Wilsonian effective action though only up to the conformal anomaly, described below.
The associated ERG equation essentially mirrors the well-known flat space ERG equation. However, even in this case there are some important differences. First of all, in order to render certain vacuum terms well defined, a set of free regulator fields is required. This Pauli-Villars regularization is reminiscent of the scheme employed in the manifestly gauge invariant ERG of~\cite{mgierg1} and also the more recent~\cite{Falls-ConformalAnomaly}, which recovers the conformal anomaly of the Maxwell theory without gauge fixing.
Secondly, the conformal anomaly must explicitly appear in the equation:
\be
	\Int{x}
	\biggl(		
		\scaling{\dfield} \, \dfield \, \fder{}{\dfield}
		- 2 g_{\mu\nu} \fder{}{g_{\mu\nu}}
		+ \cdots
	\biggr)
	e^{-{\Stot}[\dfield, g_{\alpha\beta}]}
	=
	-\CurvedInt{x}  \anomaly(x) e^{-{\Stot}[\dfield, g_{\alpha\beta}]}
,
\label{eq:Weyl-Schematic}
\ee
where, as usual, $g \equiv \det g_{\mu\nu}$ and the ellipsis contains quasi-local terms involving functional derivatives \wrt\ $\dfield$, plus some non-critical fields necessary to render the vacuum term finite. The dependence of the action on the non-critical fields is suppressed. 

The appearance of the anomaly may be understood as follows. Amongst the suppressed terms on the \lhs\ of~\eq{Weyl-Schematic} is a double functional derivative term. Acting on $\Stot$, this will generate a vacuum term. The only way this can be cancelled is by the $\Int{x} g_{\mu\nu}  \delta /\delta g_{\mu\nu}$ term. However, any vacuum terms that lie in the kernel of this operator cannot be cancelled by a local contribution to the action.%
\footnote{For local terms, $\Int{x} g_{\mu\nu}  \delta /\delta g_{\mu\nu}$ just counts the scaling dimension. Therefore, a term in the kernel of this operator cannot be cancelled by the operator acting on some other local term: either the term has the wrong scaling dimension or it is annihilated. However, this does not hold true for non-local terms, which is the origin of the Polyakov action.}
 Such vacuum terms constitute the conformal anomaly.

Now we wish to relate the conformal anomaly to the trace of the energy-momentum tensor, as defined in~\eq{T(x,y)}. To sketch this, assume that the limit $y \rightarrow x$ exists.
The expectation value of $T$ must give zero on account of the total functional derivative. Picking out the term in~\eq{T(x,y)} which resembles the classical dilatation generator and integrating over all space leads us to consider
\be
	\CurvedInt{x}
	\eval{ T(x)}
	=
	\frac{1}{\pf} \int \measure{\dfield \cdots}
	\biggl(
		\scaling{\dfield} \Int{x} \dfield \, \fder{}{\dfield}
		+ \cdots
	\biggr) e^{-\Stot[\dfield, g_{\alpha\beta}]}
	=
	0
,
\label{eq:EMT-eval}
\ee
where the non-critical regulator fields alluded to earlier are suppressed and $\pf$ is the partition function. The point about the term shown explicitly is that is may be traded using~\eq{Weyl-Schematic} to give
\be
	\CurvedInt{x} \eval{ T(x)}
	=
	\frac{1}{\pf} \int \measure{\dfield\cdots}
	\biggl[
		\Int{x}
		\biggl(
			2 g_{\mu\nu} \fder{}{g_{\mu\nu}}
			-\sqrt{g} \anomaly
		\biggr)
		+ \cdots
	\biggr] e^{-\Stot[\dfield, g_{\alpha\beta}]}
	=
	0
.
\label{eq:EMT-eval-anomaly}
\ee
As will be properly justified later, the ellipsis correspond to redundant terms---i.e.\ they can be removed by a field redefinition. Exploiting the fact that the combination of the ERG's built-in cutoff together with the regulator fields ensures ultraviolet finiteness, we pull the functional derivative through the measure (which contains an equal number of commuting and anti-commuting degrees of freedom) to yield
\be
	-2 \CurvedInt{x} g_{\mu\nu}
	\frac{1}{\sqrt{g}}\fder{W[g]}{g_{\mu\nu}}
	=
	\CurvedInt{x} \anomaly
,
\label{eq:TraceAnomaly}
\ee
where $\Schwinger \equiv -\ln \pf$. \Eqn{TraceAnomaly} is just an integrated expression relating the trace of the `quantum energy momentum tensor' to the conformal anomaly.

Demonstrating the unintegrated form using the ERG is beyond the scope of this paper: it would require considering local, as opposed to global, Weyl transformations. However, having already made contact with the flat space conformal fixed-point equation will enable us to provide some reasonable ideas on this matter, which will be done in \sect{beyond}. In the meantime, let us note that Weyl invariance in the context of the ERG has been discussed previously, most notably in~\cite{Percacci-Weyl}, which builds upon~\cite{Codello-PolyakovAction}. The latter paints a clear picture of the Polyakov action. However, the approach taken is rather different from that advocated here. In~\cite{Percacci-Weyl}, the formulation of the ERG is simply the Legendre transform of Polchinski's equation~\cite{Pol}. As such, one starts with dimensionful variables and the effective scale, $\Lambda$, explicitly appears in the effective action, regardless of whether or not it corresponds to a fixed-point.%
\footnote{In the literature on the effective average action, $k$ is frequently used instead of $\Lambda$.}
The paper subsequently explores introducing a dilaton, which is used not only to render any dimensionful couplings dimensionless but also to render the effective scale---promoted to a function of spacetime---dimensionless. By contrast, in this paper an equation is proposed for which there is no running scale and no dilaton but for which the allowed Weyl transformations are only global. Nevertheless, as we have seen, the origin of the conformal anomaly is particularly transparent. It is hoped that the approach of this paper provides a natural arena for further exploring Weyl invariance within the ERG.

The remainder of this paper is organized as follows. \Sect{ERG} starts by reviewing various essential elements of the ERG. For the purposes of illustration, the familiar example of the Gaussian fixed-point is explored; however, in contrast to earlier works, the vacuum terms contributing to $\EMT$ are carefully treated and it is shown that contributions to $\EMT(x,y)$ which potentially diverge in the $y \rightarrow x$ limit in fact cancel. Following this is a brief exploration of the various free non-critical theories which, while generally of little interest, play a star role in curved space. 

\Sect{Weyl} demonstrates how, starting from the canonical ERG, a natural generalization to curved space may be constructed.
As alluded to, divergence of a vacuum term necessitates the introduction of regulator fields which we may take to be free non-critical  theories. As an immediate application of the new equation, the Gaussian solution is presented in \sect{Gaussian_Anomaly}, with the anomaly and Polyakov action recovered for $\D=2$. Following this, in \sect{Legendre} a Legendre transform is taken, which leads to some appealing simplifications. On the one hand, the resulting equation, \eq{Legendre}, for the effective average action is structurally simpler than its Wilsonian effective action counterpart. On the other, the regulator fields disappear from the final form of this equation. It is anticipated that this equation, or the equivalent~\eq{Legendre'}, will be the starting point for future developments.
The latter generalizes in a particularly natural way to theories which do not reside at a fixed-point, as shown in~\eq{Legendre_t}.

\section{Exact RG}
\label{sec:ERG}

\subsection{Defining Equations}
\label{sec:Defining}

Throughout this section we work in $\D$-dimensional Euclidean space; the generalization to Riemannian manifolds comes later. 
Let us begin by elucidating the structure of the UV cutoff, $\cutoff$. When explicitly indicating the coordinate dependence, we shall write it as $\cutoff(x,y) = \cutoff(y,x)$. For the purposes of this paper, it is particularly convenient to define the cutoff function via a Laplace transform:
\be
	\cutoff(x,y) = 
	\LimitInt{u}{0}{\infty} \hat{\cutoff}(u)
	e^{- u (-\partial^2)}
	\DD{\D}(x-y)
.
\label{eq:cutoff-Laplace}
\ee
There is considerable freedom over the form of the kernel, $\hat{\cutoff}(u)$, though it must preserve several key properties:
\begin{enumerate}
	\item $\cutoff(x,y)$ must be quasi-local, \ie\ have a Taylor expansion in $\partial^2$;
	
	\item For large $u$, the integrand must decay at least exponentially fast (which follows from the momentum-space analysis of~\cite{Wegner-CS,Fundamentals});
	
	\item The normalization is such that 
	$\LimitInt{u}{0}{\infty} \hat{\cutoff}(u) = 1$
.
\end{enumerate}
A perfectly reasonable choice of kernel is $\hat{\cutoff}(u) = \delta(u-1)$, corresponding to an exponential cutoff function, though we are by no means restricted to this.

In order to construct the ERG equation and its special conformal partner,
we define several quantities, $G$, $G_\mu$ and $\ep$ derived from $\cutoff$:
\begin{subequations}
\begin{align}
	\bigl(
		\D + x\cdot \partial_x + y \cdot \partial_y
	\bigr)
	\cutoff (x,y)
	& = \partial^2_x G(x,y)
,
\label{eq:G}
\\
	G_\mu(x,y) 
	& = (x + y)_\mu G(x,y)
,
\\
	\ep = \ep_0 \cdot \cutoff
\label{eq:ep}
\end{align}
\end{subequations}
where, using  $\one(x,y) = \DD{\D}(x-y)$,
\be
	-\partial^2 \ep_0 = \one
\label{eq:Green}
.
\ee

It is worthwhile to consider some of these expressions in momentum space. To do so, we overload notation so that the same symbol is used for a function and its Fourier transform. This is with the understanding that if there is a single argument---usually $p$---it is interpreted as a momentum whereas if there are two arguments---usually $x,y$---they are interpreted as positions. Thus the momentum space cut-off function is simply $\cutoff(p^2)$, and decays rapidly for large $p^2$. Bearing in mind~\eq{cutoff-Laplace}, \eq{G} becomes
\be
	\der{\cutoff(p^2)}{p^2} = \hf G(p^2)
.
\label{eq:G(p^2)}
\ee
For $\D>2$, we see from~\eq{Green} that $\ep_0 = 1/p^2$. Note that the above normalization condition translates to
\be
	\cutoff(p^2 = 0) = 1
.
\label{eq:cutoff_norm}
\ee

In flat space, neglecting vacuum terms, the dilatation and special conformal generators in the ERG representation are given by~\cite{Representations}:
\begin{subequations}
\begin{align}
	\Dil 
	& = 
	\dil{\scaling{\dfield}} \dfield \cdot \fder{}{\dfield}
	+ \dfield \cdot \ep^{-1} \cdot \G \cdot \fder{}{\dfield}
	+ \hf \fder{}{\dfield} \cdot \G \cdot \fder{}{\dfield}
,
\label{eq:Dil-ERG}
\\
	\SCT_\mu
	&=
	\sct{\scaling{\dfield}}{\mu} \dfield \cdot \fder{}{\dfield}
	+ \dfield \cdot \ep^{-1} \cdot G_\mu \cdot \fder{}{\dfield}
	+ \hf \fder{}{\dfield} \cdot G_\mu \cdot \fder{}{\dfield}
	- \fp{\eta} \, \partial_\alpha \dfield \cdot \cutoff^{-1} \cdot \G \cdot \fder{}{\dfield}
\label{eq:SCT-ERG}
,
\end{align}
\end{subequations}
with the symbol $\eta$ related to the scaling dimension, $\scaling{\dfield}$, according to
\be
	\scaling{\dfield} = \frac{\D-2+\eta}{2}
.
\ee

A conformal field theory satisfies
\begin{subequations}
\begin{align}
	\Dil  e^{-\Stot} & \sim 0,
\label{eq:dil_condition}
\\
	\SCT_\mu e^{-\Stot} & = 0
\label{eq:sct_condition}
,
\end{align}
\end{subequations}
where $\sim$ indicates neglected vacuum terms. Note that translation invariance prohibits any such terms in the second constraint. Using~\eq{Dil-ERG} it is straightforward to convert the first constraint into a non-linear equation for $\Stot$; doing so yields the canonical ERG equation of~\cite{Ball}. Interestingly, this equation contains two \emph{a priori} unknown quantities: $\Stot$ and $\scaling{\dfield}$. However, imposing that solutions for $\Stot$ are quasi-local---\ie\ exhibit a derivative expansion---means that both the action and scaling dimension are self-consistently determined: the ERG equation behaves like a non-linear eigenvalue equation~\cite{TRM-Elements}. 

To conclude this section, we provide an expression for $\R$, which appears in the trace of the energy-momentum tensor~\eq{trace_EMT}. In momentum space, and for $\eta < 2$, it is given by~\cite{HO-Remarks,Fundamentals,Representations}
\be
	\R(p^2)
	=
	p^{2(\eta/2-1)} \cutoff(p^2)
	\int_0^{p^2}
	dq^2 q^{-2\eta/2} \der{}{q^2}
	\biggl[
		\frac{1}{\cutoff(q^2)}
	\biggr]
.
\label{eq:R}
\ee
In \app{Derivation} an alternative justification will be given for this equation. In the meantime, we record the following useful property:
\be
	2 p^2 \der{\R(p^2)}{p^2}
	=
	(\eta -2)\R(p^2)
	+
	 G(p^2) \cutoff^{-1}(p^2)
	\bigl(
		 p^2 \R(p^2) - 1	
	\bigr)
.
\label{eq:dR/dp^2}
\ee

\subsection{Example}
\label{sec:Example}

Having introduced the basic formalism, we now illustrate some of the themes discussed above.
To this end, let us examine the Gaussian fixed-point, for which $\scaling{\dfield} = (\D-2)/2$. The Wilsonian effective action exists as a curve in theory space, parametrized by $-\infty < b < 1$:
\be
	\SGauss_b[\dfield]
	\sim
	\hf
	\dfield \cdot \ep^{-1}_b \cdot \dfield
\label{eq:Gaussian_Flat}
\ee
where, for now, we have dropped vacuum terms and
\be
	\ep^{-1}_b
	=
	\ep^{-1}
	\cdot
	\bigl(
		\one-b \cutoff
	\bigr)^{-1}
.
\label{eq:Gaussian_TP_Flat}
\ee
This solution can be checked using~\eqs{Dil-ERG}{dil_condition}, though it will anyway be derived in the subsequent analysis on curved space.
A useful piece of intuition is provided by observing that, for small $p^2$, $\ep^{-1}_b(p^2) \sim p^2/(1-b)$: in essence, the parameter $b$ carries the freedom to rescale the field. Indeed, by coupling a source, it is straightforward to show that the two-point correlation function is given by $(1-b)\vert x - y\vert^{-(\D-2)}$~\cite{Fundamentals}.

Ordinarily, the non-critical fixed-point occurring at $b = 1$ is of little interest. However, in this paper, it will play an important role. Note that, on account of the normalization condition~\eq{cutoff_norm}, the action starts not at $\order{\partial^2}$ but has a mass-like contribution for $b=1$. Furthermore, the two-point correlation function vanishes: there are no long-range correlations hence the moniker `non-critical'. As will be seen later, we shall utilize certain non-critical fields to regularize a divergent vacuum term.

We now construct $\EMT_b$ using~\eq{T(x,y)}. Since the anomalous dimension, $\eta$, is zero, \eq{R} reduces to
\be
	\R = \ep_0 - \ep = \ep_0 \cdot(\one - \cutoff)
,
\label{eq:R_Gauss}
\ee
whereupon we find that, as alluded to earlier, the potentially divergent vacuum terms cancel:
\be
	\EMT_b(x,y) = 
	\frac{(\D-2)(1-b)}{2}
	\biggl[
		\Bigl(
		\dfield \cdot 
		\bigl(
			\one-b \cutoff
		\bigr)^{-1}
		\Bigr)(x)
		\Bigl(
		\bigl(
			\one-b \cutoff
		\bigr)^{-1}
		\cdot
		\partial^2 \dfield
		\Bigr)(y)
		+
		\Bigl(
		\cutoff \cdot 
		\bigl(
			\one-b \cutoff
		\bigr)^{-1}
		\Bigr)
		(x,y)
	\biggr]
.
\ee
The short-distance limit can therefore be safely taken 
with the vacuum term readily evaluated in momentum space:
\be
	\Bigl(
		\cutoff \cdot 
		\bigl(
			\one-b \cutoff
		\bigr)^{-1}
	\Bigr)
	(x,x)
	=
	\MomInt{p} \frac{\cutoff(p^2)}{1 - b \cutoff(p^2)}
.
\ee
This is finite, courtesy of the UV cutoff function.
Notice that, for the non-critical fixed-point at $b=1$, $\EMT$ vanishes. Furthermore, for the canonically normalized Gaussian fixed-point with $b=0$, we have

\be
	\EMT_0(x) \sim
	\frac{\D-2}{2}  
	\partial^2 \dfield
	\times \dfield
,
\ee
as expected.

It may be readily checked that $\Int{x} \EMT_b(x)$ is tangent to the Gaussian fixed-point action, providing a concrete example of the earlier claim that the trace of the energy-momentum tensor may be identified with the exactly marginal, redundant perturbation which generates motion along a line of physically equivalent fixed-points.

We conclude this section by discussing the vacuum contribution to $\Stot$, using its structure to anticipate the additional  regularization that will supplement the ERG upon its generalization to curved space.
Formally, acting on $\exp -\Stot^\mathrm{Gauss}_b$ (with the action given by~\eq{Gaussian_Flat}) the generator~\eqref{eq:Dil-ERG} generates a momentum space vacuum term
\be
	(2\pi)^{\D} \DD{\D}(0)
	\hf
	\MomInt{p}
	\bigl(-\ep^{-1}_b(p^2) G(p^2)\bigr)
.
\ee
In curved space, the delta-function will be replaced by the volume of the space and so, for now, we will focus just on the integral. Our strategy for regularizing this will be to add one bosonic field together with an anticommuting scalar and its conjugate, ensuing the number of commuting/anti-commuting degrees of freedom are the same. The action for all of these additional fields will be chosen to be the non-critical Gaussian theory---i.e.\ with $b=1$. The effect on the integrand of the vacuum term is to replace it with
\be
	-\hf
	\bigl(
		\ep^{-1}_b(p^2) - \ep^{-1}_1(p^2)
	\bigr)
	G(p^2)
	=
	\frac{1-b}{2}
	\frac{G(p^2) p^2}{(1- b \cutoff(p^2))(1 - \cutoff(p^2))}
	=
	-p^2 \der{}{p^2}
	\ln \frac{1- \cutoff(p^2)}{1- b \cutoff(p^2)}
.
\label{eq:regularized_vacuum_flat}
\ee
Notwithstanding the divergent pre-factor, the momentum integral is now convergent. In the UV, regularization is provided by $G(p^2)$, which decays at least exponentially fast, with the denominator being UV safe. On the other hand, in the IR,
\be
	\lim_{p^2\rightarrow 0}
	\frac{1-b}{2}
	\frac{G(p^2) p^2}{(1- b \cutoff(p^2))(1 - \cutoff(p^2))}
	=
	-1
	+ \order{p^2}
,
\label{eq:Universal}
\ee
on account of~\eq{cutoff_norm}.
As will be seen later, the universality of the leading coefficient---it is independent both of $b$ and the cutoff function---is essential for the recovery of the conformal anomaly.

As a concrete example of the convergence of the momentum integral, let us work in $\D=2$ and take $\cutoff(p^2) = e^{-p^2}$. Integrating by parts and noting that the surface terms vanish, the integral can be expressed in terms of dilogarithms:
\be
	\int_0^\infty dp^2 p^{2}
	\der{}{p^2}
	\ln \frac{1- e^{-p^2}}{1- b e^{-p^2}}
	=
	\mathrm{Li}_2(1) - \mathrm{Li}_2(b)
.
\ee

\subsection{The Non-Critical Fixed-Points}
\label{sec:NonCritical}

As emphasised already, non-critical fixed-points will have an important role to play. We have already seen the non-critical theory arising from the $b \rightarrow 1$ limit of the Gaussian fixed-point; now we examine the complete family of free, non-critical theories. To this end, we look for two-point solutions to 
the ERG equation of the form:
\be
	\Stot[\dfield] \sim \hf \dfield \cdot f \cdot \dfield
,
\label{eq:S-noncritical}
\ee
for some quasi-local $f$. Recall that the ERG equation follows from allowing~\eq{Dil-ERG} to act on $\exp - \Stot$. Working in momentum space yields
\be
	\biggl(
		2 p^2 \der{}{p^2} + \eta - 2 + 2 G(p^2) \ep^{-1}(p^2)
	\biggr) f(p^2)
	-f^2(p^2) G(p^2)
	=
	0
.
\ee
This may be linearized by rewriting in terms of $1/f$. It is then straightforward to check, with the help of~\eqs{R}{dR/dp^2}, that
\be
	f(p^2) = \frac{\ep^{-1}(p^2)}{p^2 \R(p^2) + (1-b)\cutoff(p^2) p^{2\eta/2}}
,
\ee
with appropriate restrictions on $\eta$ and/or $b$ to ensure quasi-locality.
As a simple check note that for $\eta = 0$ we have, recalling~\eq{R_Gauss}, $p^2\R(p^2) = 1 - \cutoff(p^2)$ and the solution reduces to the Gaussian one~\eq{Gaussian_Flat}. In addition to the Gaussian fixed-point, there is a family of critical fixed-points with $\eta = -2, -4, \ldots$, though these are non-unitary in Minkowski space. However, our real interest is in the family of non-critical fixed-points with $b=1$. These take the form
\be
	f = \cutoff^{-1} \cdot \sigma
,
\label{eq:NonCritical_TP}
\ee
where
\be
	\Rinv \equiv \R^{-1}
.
\label{eq:Rinv}
\ee
Notice that, by Taylor expanding in $p^2$, $\sigma$ behaves like a mass term in the IR, confirming the non-critical nature of these fixed-points, which exist for all $\eta <2$.

\section{An ERG equation for Curved Space}
\label{sec:Weyl}

In this section, a generalization of the canonical ERG equation to curved space is proposed. The basic idea is presented in \sect{Weyl-Initial}. \Sect{Weyl-Cutoff} is devoted to understanding how the cutoff function behaves under \emph{local} Weyl transformations, confirming the difficulty of dealing with these within the ERG. \Sect{Weyl-Full} proposes a curved space equation which fully regularizes the vacuum contributions and reduces to the canonical ERG equation on flat space. \Sect{EMT} fleshes out the identification, sketched in the introduction, of the anomaly appearing in the new ERG equation to the conventional definition in terms of the trace of the quantum energy-momentum tensor.

\subsection{Initial Considerations}
\label{sec:Weyl-Initial}

To facilitate working in curved space, it will pay to tweak our notation. Given a field, $\psi(x)$, define
\be
	\fderhat{}{\psi(x)} \equiv \frac{1}{\sqrt{g}} \fder{}{\psi(x)}
\ee
and, in a similar vein,
\be
	\one(x,y) = \frac{1}{\sqrt{g}} \DD{\D}(x-y)
.
\label{eq:one}
\ee
Next, we redefine the dot notation introduced in~\eq{dot} such that, for example,
\be
	\bigl(
		\cutoff^{-1} \cdot \dfield
	\bigr)(x)
	=
	\CurvedInt{y}
	\cutoff^{-1}(x,y) \, \dfield(y)
	,
	\qquad
	\dfield \cdot \dfield = \CurvedInt{x} \dfield(x) \dfield(x)
.
\label{eq:dot_notation-Curved}
\ee
Finally, we shall adopt the convention that rather than simply replacing $\partial^2 \rightarrow \nabla^2$, we instead make the replacement
\be
	-\partial^2 \rightarrow \WeylLaplacian  = -\nabla^2 + \frac{\D-2}{4(\D-1)} \Ricci
,
\label{eq:WeylLaplacian} 
\ee
where $\Ricci$ is the Ricci scalar. It is not strictly necessary to do this, but doing so simplifies various equations and calculations if we do so. 
The motivation for this is that, classically, the Weyl-invariant Gaussian theory has action
\be
	\classical{\Stot}[\dfield]
	=
	\CurvedInt{x}
	\dfield
	\WeylLaplacian
	\dfield
.
\label{eq:GaussianCurvedTP}
\ee

The basic strategy for generalizing the ERG to curved space has been outlined in the introduction; now we fill in the gaps. The starting point is the classical statement of Weyl invariance, \eq{Weyl-cl}. Our first task is to understand how the statement of conformal invariance given by~\eq{classical_conformal} is recovered in the flat space limit, for which we follow~\cite{HO-CFTLectureNotes}.

Given a vector field, $v_\mu(x)$, we may construct the Lie derivative, $\Lie{v}$. Demanding invariance under infinitesimal diffeomorphisms yields
\be
	\Int{x}
	\biggl(
		\bigl(
			\nabla_\mu v_\nu + \nabla_\nu v_\mu
		\bigr)
		\fder{}{g_{\mu\nu}}
		+
		\Lie{v} \dfield \fder{}{\dfield}
	\biggr)
	e^{-\classical{\Stot}[\dfield, g_{\alpha\beta}]}
	= 0
.
\label{eq:diffeo}
\ee
Conformal Killing vectors satisfy
\be
	\nabla_\mu v_\nu + \nabla_\nu v_\mu
	= 2 \weyl g_{\mu\nu}
.
\label{eq:Killing}
\ee
In the flat space limit this has solutions corresponding to scale and special conformal transformations:
\be
	v_\mu^{\mathrm{scale}} = \kappa x_\mu,
	\qquad
	v_\mu^{\mathrm{special}} = b_\mu x^2 
	- 2 x_\mu b \cdot x
,
\ee
for infinitesimal parameters, $\kappa$ and $b_\mu$, which are independent of $x$.
In both cases~\eq{Killing} implies that $\partial_\mu v^\mu = \D \weyl$ and so
\be
	\weyl^{\mathrm{scale}} = \kappa,
	\qquad
	\weyl^{\mathrm{special}}
	=
	-2 b \cdot x
.
\label{eq:weyl-specific}
\ee

Working in the flat space limit, substitute~\eq{Killing} into~\eq{diffeo} and employ~\eq{Weyl-cl}
to give
\begin{subequations}
\begin{align}
	\Int{x}
	\biggl(		
		\weyl \scaling{\dfield} \, \dfield \, \fder{}{\dfield}
		+
		\Lie{v} \dfield \fder{}{\dfield}
	\biggr)
	e^{-\classical{\Stot}[\dfield, g_{\alpha\beta}]}
	&
	= 0
,
\label{eq:Weyl-Killing}
\\
	\Int{x}
	\weyl
	g_{\mu \nu}
	\fder{}{g_{\mu\nu}}
	e^{-V[g_{\alpha \beta}]}
	& = 0
,
\end{align}
\end{subequations}
where $V$ comprises those contributions to the action which are independent of $\dfield$.

Recalling that, on scalars, the Lie derivative reduces to the directional derivative, $v \cdot \partial$, it is straightforward to check that, on flat space,
\eq{Weyl-Killing} does indeed encode classical conformal invariance. For example, consideration of scale transformations gives
\be
	\kappa
	\Int{x}
	\biggl(		
		\scaling{\dfield} \, \dfield \, \fder{}{\dfield}
		+
		x \cdot \partial \dfield \fder{}{\dfield}
	\biggr)
	e^{-\classical{\Stot}[\dfield]}
	= 0
	\qquad
	\Rightarrow
	\qquad
	\dil{\scaling{\dfield}}
	\dfield \cdot \fder{}{\dfield}
	e^{-\classical{\Stot}[\dfield]}
	= 0
.
\ee

The strategy for generalizing the ERG equation to curved space should now be clear: we seek a generalization of~\eq{Weyl-cl} that, for constant $\omega$, reduces to~\eq{dil_condition} on flat space. For conformal theories, it must be that~\eq{sct_condition} is recovered in flat space; we revisit this in \sect{beyond}.
 
\subsection{The Cutoff Function}
\label{sec:Weyl-Cutoff}

Before proceeding any further, a thorough understanding of the cutoff function and, in particular, its response to local Weyl transformations is desirable. As a warm up, it is worth explicitly demonstrating the Weyl invariance of the classical Gaussian action in curved space, \eq{GaussianCurvedTP}. 
To confirm this, note that under an infinitesimal Weyl transformation,
\be
	g_{\mu\nu} \rightarrow g_{\mu\nu}(1+2\weyl),
	\quad
	g^{\mu\nu} \rightarrow g^{\mu\nu}(1-2\weyl),
	\quad
	\sqrt{g} \rightarrow \sqrt{g}(1+\D\weyl),
	\quad
	R \rightarrow (1-2\weyl) R - 2(\D-1) \nabla^2 \weyl
.
\label{eq:weyl-g}
\ee
From this is follows that
\be
	\WeylLaplacian 
	\rightarrow \WeylLaplacian + \delta_\weyl \WeylLaplacian + \order{\weyl^2}
	=
	(1-2 \weyl) \WeylLaplacian
	-\frac{\D-2}{2}
	\Bigl(
		2\partial_\mu \weyl \, g^{\mu\nu} \partial_\nu + \nabla^2 \weyl
	\Bigr)
,
\label{eq:delta_wDelta_W}
\ee
allowing us to deduce that
\begin{subequations}
\begin{align}
	\delta_\weyl \WeylLaplacian 
	& = \frac{\D-2}{2} \WeylLaplacian \weyl
	- \frac{\D+2}{2} \weyl \WeylLaplacian
\\
	& = - 2\weyl \WeylLaplacian 
	+ \frac{\D-2}{2} 
	\bigl[
		\WeylLaplacian, \weyl
	\bigr]
.
\label{eq:delta_weyl}
\end{align}
\end{subequations}
Given that the field transforms as
\be
	\dfield \rightarrow \dfield
	\bigl(1 - \scaling{\dfield} \, \weyl \bigr)
\label{eq:weyl-phi}
\ee
invariance of $\classical{\Stot}[\dfield]$ follows if $\dfield$ has scaling dimension $(\D-2)/2$.

Moving on to the cutoff function, we overload notation such that, for example,
\be
	\cutoff(x,y) = \cutoff(\WeylLaplacian) \one(x-y)
.
\label{eq:K(x,y)-Curved}
\ee
As in~\eq{cutoff-Laplace} we utilize a Laplace transform
\be
	\cutoff(\WeylLaplacian) = \LimitInt{u}{0}{\infty} \hat{\cutoff}(u) e^{-u \WeylLaplacian}
,
\label{eq:cutoff(WeylL)-Laplace}
\ee
the point of which is that the exponential has a simple response to infinitesimal Weyl transformations:
\be
	\delta_\weyl e^{-u \WeylLaplacian} =
	-u \LimitInt{a}{0}{1} e^{-a u \WeylLaplacian} 
	\bigl(\delta_\weyl \WeylLaplacian \bigr)
	e^{-(1-a) u \WeylLaplacian} 
,
\label{eq:cutoff-response}
\ee
with $\delta_\weyl \WeylLaplacian $ given~\eq{delta_weyl}. \Eqn{cutoff-response} may be checked by expanding the exponentials and employing
\be
	\LimitInt{a}{0}{1} a^n (1-a)^m = \frac{n! m!}{(1+m+n)!}
.
\ee
From~\eq{cutoff-response} it follows that
\be
	\delta_\weyl \cutoff(\WeylLaplacian)
	= 
	-\LimitInt{u}{0}{\infty} 	
	u \hat{\cutoff}(u)
	\LimitInt{a}{0}{1} e^{-a u \WeylLaplacian} 
	\bigl(\delta_\weyl \WeylLaplacian \bigr)
	e^{-(1-a) u \WeylLaplacian} 
.
\ee

A piece of intuition which may be useful follows from taking $\weyl$ to be unity in~\eq{cutoff-response} and considering some $F(\WeylLaplacian)$. On the one hand,
\be
	\delta_\weyl F(\WeylLaplacian) 
	\Bigr \vert_{\weyl = 1}
	=
	-2
	\LimitInt{u}{0}{\infty}
	\hat{F}(u) u \der{}{u} e^{-u\WeylLaplacian}
	=
	2
	\LimitInt{u}{0}{\infty}
	e^{-u\WeylLaplacian}
	\der{}{u} u \hat{F}(u)
.
\label{eq:intuition}
\ee
On the other hand, for some parameter, $z$,
\be
	z \der{}{z} F(z)
	=
	\LimitInt{u}{0}{\infty}
	\hat{F}(u)
	u
	\der{}{u}
	e^{-uz}
	=
	-
	\LimitInt{u}{0}{\infty}	
	e^{-uz}
	\der{}{u}
	u
	\hat{F}(u)
.
\label{eq:zd/dz}
\ee
In flat space we may take $z = p^2$ and so, roughly speaking, can map between curved space and flat space with the mnemonic $\delta_\weyl \leftrightarrow -p \cdot \partial_p$. Indeed, for global Weyl transformations, writing down a curved space ERG equation largely follows a simple recipe: make the replacement~\eq{WeylLaplacian}  and insert factors of $\sqrt{g}$ in the appropriate places. However, for local Weyl transformations such a simple recipe cannot work, on account of~\eq{cutoff-response}. This is reminiscent of the challenges of constructing a manifestly gauge invariant ERG~\cite{mgierg1,Fundamentals}.

To conclude this section, recall~\eq{K(x,y)-Curved} and observe that
\be
	\delta_\weyl F(x, y) 
	\Bigr \vert_{\weyl = 1}
	=
	\delta_\weyl F(\WeylLaplacian) 
	\Bigr \vert_{\weyl = 1}
	\one(x-y)
	-\D F(x, y)
\label{eq:Full-delta_weyl}
\ee

\subsection{The Equation}
\label{sec:Weyl-Full}

We have accepted that our goal is a generalization of the ERG to curved space which encodes invariance only under global Weyl transformations. However, the simple recipe just described misses two details: one is the necessity of regularizing the vacuum term and the other is the emergence of the conformal anomaly. With these points in mind, let us introduce
\begin{multline}
	\Dil^{\Weyl}(x,y)
	=
	\biggl(	
		\scaling{\dfield} \, \dfield \times \fderhat{}{\dfield}
		- 2 g_{\mu\nu} \times \fderhat{}{g_{\mu\nu}}
		+
		\hf\kernel{\dfield \cdot \ep^{-1}}{\fderhat{}{\dfield}}	
	\biggr)(x)
\\
	+
	\biggl(
		\frac{1}{4}\pskernel{\fderhat{}{\dfield}}{\fderhat{}{\dfield}}
	\biggr)(x,y)
	+
	\cdots
,
\label{eq:Dil-Weyl}
\end{multline}
where the ellipsis indicates the anticipated regulator terms and for $A(x)$, $B(x)$, we understand $(A \otimes B)(x,y) = A(x) B(y)$. 

The problem with the vacuum terms may be illustrated by supposing that the flat space Gaussian solution directly generalizes to curved space (this will be justified, in the next section). Recalling~\eq{Gaussian_Flat}, we therefore expect to find that
\be	
	e^{\Stot^{\mathrm{Gauss}}_b}
	\Dil_{\Weyl}(x,y) e^{-\Stot^{\mathrm{Gauss}}_b}
	=
	-\hf
	\bigl(
		G \cdot \ep^{-1}_b
	\bigr)(x,y)
.
\ee
The reason for point-splitting is now apparent: the limit $y \rightarrow x$ does not exist. Intuitively, this can be seen by returning to flat space and recalling the discussion of vacuum terms in \sect{Example}.

Point splitting at least allows us to regularize the divergence. But we now need to cancel the potentially divergent term before taking the short-distance limit, and herein lies the problem. The natural way to cancel this vacuum term is via a vacuum contribution to the action which, when hit by $-2 g_{\mu\nu} \cdot \hat{\delta} / \delta g_{\mu\nu}$, generates the necessary term. The difficulty is particularly transparent for the Gaussian solution with $b=0$, for which the desired vacuum term is, formally,
\[
	\Int{x} \sqrt{g} \ln \cutoff
.
\]
However, this itself is divergent! Again, intuition for this can be found in flat space. Recalling that the cutoff function decays at least exponentially, this would correspond, in the best case, to something which goes like $\MomInt{p} p^2$. Therefore, while point splitting is a necessary step, it is not sufficient to provide us with a good curved space generalization of the ERG.

The solution to the problem of the vacuum terms is, as anticipated earlier, to introduce a set of regulator fields comprising a bosonic field plus an anti-commuting scalar and its conjugate. In preparation for this, let us rewrite~\eq{Dil-Weyl}
as follows:
\be
	\Dil_\Weyl(x,y)
	=
	\biggl(
		\linear{\dfield}
		-2g_{\mu\nu} \times \fderhat{}{g_{\mu\nu}}
	\biggr)(x)
	+\quadratic{\dfield}{\dfield}(x,y)
,
\ee
with
\begin{subequations}
\begin{align}
	\linear{\dfield}
	& \equiv
	\scaling{\dfield} \, \dfield \times \fderhat{}{\dfield}
	+
	\hf\kernel{\dfield \cdot \ep^{-1}}{\fderhat{}{\dfield}}
\label{eq:C_phi}	
\\
	\quadratic{\dfield}{\psi}
	& \equiv
	\frac{1}{4} \pskernel{\fderhat{}{\dfield}}{\fderhat{}{\psi}}
.
\end{align}
\end{subequations}

Denoting the regularizing scalar field by $\field$ and the anti-commuting fields by $\chi$, $\overline{\chi}$, we take
\be
	\Dil_\Weyl(x,y)
	=
	\biggl(
		\linear{\dfield}
		+\linear{\field}
		+\linear{\chi}
		+\linear{\overline{\chi}}
		-2g_{\mu\nu} \times \fderhat{}{g_{\mu\nu}}
	\biggr)(x)
	+
	\biggl(
		\quadratic{\dfield}{\dfield}
		+\quadratic{\field}{\field}
		-2\quadratic{\chi}{\overline{\chi}}
	\biggr)(x,y)
\ee
and seek solutions such that
\be
	\CurvedInt{x}
	\Bigl(
		\lim_{y \rightarrow x}
		\Dil_\Weyl(x,y)
		+ \anomaly(x)
	\Bigr)
	e^{-\Stot[\dfield, \field, \chi, \overline{\chi}]}
	=
	0.
\label{eq:Weyl-ERG}
\ee
We have constructed this equation with the premise that $\dfield$ is physical, with all other fields present to provide regularization. Indeed, let us emphasise that the only necessary regularization is for the vacuum term and, consequently, the various regulator fields are decoupled both from each other and from the physical field. Furthermore, given a critical fixed-point, it will always be sufficient to take these regulator fields to be at the free, non-critical fixed-point with the corresponding value of $\eta$ (recall \sect{NonCritical}). This will be demonstrated in \sect{Legendre}.

\subsection{The Trace of the Energy-Momentum Tensor}
\label{sec:EMT}

In this section, we properly justify the argument presented in the introduction that the anomaly appearing in the ERG equation coincides with the standard definition. To achieve this, we start by generalizing~\eq{T(x,y)} to curved space:
\begin{multline}
	\EMT(x, y) e^{-\Stot[\dfield, \field, \chi, \overline{\chi}]}
	=	
	\scaling{\dfield}
	\biggl\{
		\biggl(
			\fderhat{}{\dfield} \cdot \cutoff
		\biggr)(x)
		\biggl(
			\cutoff^{-1} \cdot \dfield
			+ \R \cdot \fderhat{}{\dfield}
		\biggr)(y)
		+ \dfield \leftrightarrow \field
	\\
		-
		\biggl(
			\fderhat{}{\chi} \cdot \cutoff
		\biggr)(x)
		\biggl(
			\cutoff^{-1} \cdot \chi
			+ \R \cdot \fderhat{}{\overline{\chi}}
		\biggr)(y)
		- \chi \leftrightarrow \overline{\chi}
	\biggr\}
	e^{-\Stot[\dfield,\field, \chi, \overline{\chi}]}
.
\end{multline}
From this expression it is clear that the combination of physical and regulator fields ensures the cancellation of certain vacuum terms:
\begin{multline*}
	\biggl\{
		\biggl(
			\fderhat{}{\dfield} \cdot \cutoff
		\biggr)(x)
		\bigl(
			\cutoff^{-1} \cdot \dfield
		\bigr)(y)
		+
		\dfield \leftrightarrow \field
		-
		\biggl(
			\fderhat{}{\chi} \cdot \cutoff
		\biggr)(x)
		\bigl(
			\cutoff^{-1} \cdot \chi
		\bigr)(y)
		+ \chi \leftrightarrow \overline{\chi}		
	\biggr\}
	e^{-\Stot[\dfield, \field, \chi, \overline{\chi}]}
\\
	=
	\biggl\{	
		\bigl(
			\cutoff^{-1} \cdot \dfield
		\bigr)(y)
		\biggl(
			\fderhat{}{\dfield} \cdot \cutoff
		\biggr)(x)
		+
		\dfield \leftrightarrow \field
		+
		\bigl(
			\cutoff^{-1} \cdot \chi
		\bigr)(y)
		\biggl(
			\fderhat{}{\chi} \cdot \cutoff
		\biggr)(x)
		+ \chi \leftrightarrow \overline{\chi}	
	\biggr\}
	e^{-\Stot[\dfield, \field, \chi, \overline{\chi}]}
.
\end{multline*}
The \rhs\ has no vacuum divergences, so we may take the limit $y \rightarrow x$ and integrate over all space, whereupon the resulting terms may be traded using the ERG equation~\eq{Weyl-ERG}. Considering the expectation value of the integral of $T(x)$---as in~\eq{EMT-eval}---all terms with two functional derivatives are in fact total functional derivatives and so vanish. Amongst the survivors are terms like the final contribution to $\linear{\dfield}$ in~\eq{C_phi}. Summing over the physical and regulator fields, these may be converted into total functional derivative terms since the vacuum terms cancel. The disappearance of the various total derivative terms justifies discarding the ellipsis in~\eq{EMT-eval-anomaly}, confirming~\eq{TraceAnomaly}.

\section{The Gaussian Anomaly}
\label{sec:Gaussian_Anomaly}

\subsection{Physical Field Dependence}

Before giving the complete Gaussian solution---from which the anomaly may be extracted---we focus on the physical field dependence and so take
\be
	\SGauss_b[g_{\mu\nu}, \dfield]
	=
	\hf
	\dfield \cdot \ep^{-1}_b \cdot \dfield
	+ \cdots
,
\label{eq:curved_gaussian_ansatz}
\ee
where the ellipsis represents both the regulator and vacuum pieces.
Intuitively, we anticipate that the curved space form of $\ep^{-1}_b$ follows from making the replacement~\eq{WeylLaplacian} in the flat space version. However, there is no need to assume this: it will be independently verified in this section. For simplicity, though, we will take $\eta = 0$. It is straightforward to amend the analysis below to show that this is enforced as a consistency condition; alternatively, it follows by inspection in \sect{Legendre}.

Now consider the effect of an infinitesimal Weyl transformation,
which may be deduced from~\eqs{weyl-g}{weyl-phi}. For brevity, we introduce
\be
	\field_b \equiv \ep^{-1}_b \cdot \dfield
\label{eq:field_b}
\ee
in terms of which we have:
\begin{align}
	\delta_\weyl \hf \dfield \cdot \ep^{-1}_b \cdot \dfield
	& =
	\delta_\weyl \hf \CurvedInt{x} \dfield(x) \bigl(\ep^{-1}_b\cdot \dfield\bigr)(x)
\nonumber
\\
	& = 
	\biggl( \D - 2\frac{\D-2}{2} \biggr)
	\hf \CurvedInt{x} \weyl(x) \dfield(x) \field_b(x)
	-\hf \CurvedInt{x} \field_b(x)
	\bigl(\delta_\weyl \ep_b(\WeylLaplacian) \bigr)
	\field_b(x)
.
\end{align}
Exploiting~\eq{intuition} to process the final term yields, after some simple manipulations
\be
	\delta_\weyl \hf \dfield \cdot \ep^{-1}_b \cdot \dfield \biggr\vert_{\weyl = 1}
	=
	-
	\CurvedInt{x} \LimitInt{u}{0}{\infty}
	\field_b(x)
	u \der{\hat{\ep}_b(u)}{u} 
	\field_b(x)
.
\label{eq:Gaussian-first}
\ee
To check that we are on the right track, consider the flat space relationship
\be
	\hf
	G(p^2) = 
	\biggl(
		p^2 \der{}{p^2} + 1
	\biggr)
	\ep(p^2)
,
\label{eq:G-from-ep}
\ee
from which it follows that
\be
	\hf \hat{G}(u) = - u \der{\hat{\ep}(u)}{u}
.
\label{eq:G-Laplace}
\ee
Taking $b=0$ for the purposes of illustration $b=0$, the \rhs\ simply reduces to $\hf\dfield \cdot \ep^{-1} \cdot \G \cdot \ep^{-1} \cdot \dfield$, which is what we would expect in flat space.

Moving to the next term on the \rhs\ of~\eq{Dil-Weyl} leads us to consider
\be
	\dfield \cdot \ep^{-1}
	\cdot
	G
	\cdot
	\fderhat{}{\dfield}
	e^{-\hf \dfield \cdot \ep^{-1}_b \cdot \dfield}
\\
	=
	-
	e^{-\hf \dfield \cdot \ep^{-1}_b \cdot \dfield}	
	\field_0
	\cdot \G \cdot 
	\field_b
,
\label{eq:Gaussian-second}
\ee
whereas the third term in~\eq{Dil-Weyl} yields, in the short-distance limit and up to a vacuum term
\be
	\frac{1}{2}
	\fderhat{}{\dfield}
	\cdot
	G
	\cdot
	\fderhat{}{\dfield}
	e^{-\hf \dfield \cdot \ep^{-1}_b \cdot \dfield}
	\sim
	e^{-\hf \dfield \cdot \ep^{-1}_b \cdot \dfield}
	\frac{1}{2}
	\field_b \cdot  \G \cdot \field_b
.
\label{eq:Gaussian-third}
\ee

Combining~\eqss{Gaussian-first}{Gaussian-second}{Gaussian-third} yields
 \be
 	\LimitInt{u}{0}{\infty}
 	\biggl(
	 	\ep^{-1}_b
		 u \der{\hat{\ep}_b(u)}{u} 
		+ \bigl(
			 \ep^{-1}
			 -\smallhf \ep^{-1}_b
		\bigr)
		\hat{G}(u)
	\biggr)
	e^{-u \WeylLaplacian}
	= 0
.
 \ee
If we choose the ERG kernel, $G$, according to~\eq{G-Laplace} then this equation is solved by~\eq{Gaussian_TP_Flat}, precisely as expected. For $b=0$ this follows from inspection; in the general case this may be fleshed out as follows. First observe that
\be
	\ep^{-1} 
	-\hf \ep^{-1}_b
	=
	\hf \ep^{-1}_b \cdot 
	\bigl(\one - 2b\cutoff \bigr)
,
\ee
which leads us to consider
\[
	\LimitInt{u}{0}{\infty}
	\biggl(
		u \der{\hat{\ep}_b(u)}{u}
		+
		\hf \bigl(\one - 2b\cutoff(\WeylLaplacian) \bigr)
		\hat{G}(u)
	\biggr)
	e^{-u \WeylLaplacian}
.
\]
This can be seen to vanish upon using the flat space relationship
\be
	\biggl(
		p^2 \der{}{p^2} + 1
	\biggr)
	\ep_b(p^2)
	=
	\smallhf \bigl(1 - b \cutoff(p^2)\bigr) G(p^2) 
	- \smallhf b \cutoff(p^2) G(p^2)
	=
	\smallhf \bigl(1 - 2 b \cutoff(p^2)\bigr) G(p^2)
,
\ee
together with $-u \, d / du \leftrightarrow p^2 d / dp^2 + 1$.

\subsection{The Anomaly in $\D=2$}

In this section we present the full Gaussian solution to~\eq{Weyl-ERG} and show how, in $\D=2$, the conformal anomaly may be recovered.  Adding in the regulator terms to~\eq{curved_gaussian_ansatz}
we have, up to vacuum terms:
\be
	\Stot^{\mathrm{Gauss}}_b[\dfield, \field, \chi, \overline{\chi}, g_{\mu\nu}]
	\sim
	\hf
	\dfield \cdot \ep^{-1}_b \cdot \dfield
	+
	\hf
	\field \cdot \ep^{-1}_1 \cdot \field
	+
	\overline{\chi} \cdot \ep^{-1}_1 \cdot \chi
,
\label{eq:Gassian_Curved}
\ee
where we emphasise that the regulator fields have been chosen to be at their non-critical fixed-points. Now consider 
the vacuum contribution generated by the various double functional derivative terms:
\be
	-\frac{1}{4}
	\pskernel{\fderhat{}{\dfield}}{
		\fderhat{\Stot^{\mathrm{Gauss}}_b}{\dfield}}
	+ \cdots
	=
	\frac{1-b}{2}
	\Bigl(
	\ep_0^{-1} \cdot 
	\bigl(
		\one - b \cutoff
	\bigr)^{-1}
	\cdot
	\bigl(
		\one - \cutoff
	\bigr)^{-1}
	\cdot G
	\Bigr)(x,y)
,
\label{eq:GFP-GeneratedVac}
\ee
where the ellipsis represents the regulator contributions. 
To make progress we must  understand the limit $y \rightarrow x$. First rewrite the \rhs\ as the Laplace transform of a kernel, $F$:
\be
	\frac{b-1}{2}
	\ep_0^{-1} \cdot 
	\bigl(
		\one - b \cutoff
	\bigr)^{-1}
	\cdot
	\bigl(
		\one - \cutoff
	\bigr)^{-1}
	\cdot G
	=
	\LimitInt{u}{0}{\infty} \hat{F}(u) e^{-u \WeylLaplacian} \one
.
\label{eq:Laplace_transform}
\ee
Observe a crucial property: expanding in powers of $\WeylLaplacian$, the leading coefficient is universal---i.e.\ independent of the details of the cutoff function and the value of $b$:
\be
	\frac{b-1}{2}
	\ep_0^{-1} \cdot 
	\bigl(
		\one - b \cutoff
	\bigr)^{-1}
	\cdot
	\bigl(
		\one - \cutoff
	\bigr)^{-1}
	\cdot G
	=
	\one
	+ \order{\WeylLaplacian}
.
\ee
We saw the flat space manifestation of this in~\eq{Universal}. Together with~\eq{Laplace_transform}, this implies that
\be
	\int_0^\infty ds \hat{F}(s) = 1.
\label{eq:Universal-F}
\ee
For the action to solve the ERG equation, the vacuum terms must satisfy
\be
	2 g_{\mu\nu} \cdot \fderhat{}{g_{\mu\nu}} 
	V[\dfield, g_{\alpha\beta}]
	=
	\Tr 
	\biggl(
		\LimitInt{u}{0}{\infty} \hat{F}(u) e^{-u \WeylLaplacian} \one
	\biggr)
	- 
	\Tr \anomaly
.
\label{eq:VAC}
\ee
Since the final term lies in the kernel of $\Int{x} g_{\mu\nu}  \delta /\delta g_{\mu\nu}$, to compute $\anomaly$
we simply isolate the contribution to the preceding term against which it cancels. Focussing on $\D=2$, where we recall that $\WeylLaplacian = - \nabla^2$, and
using the heat-kernel expansion yields
\be
	\Tr 
	\biggl[
		\LimitInt{u}{0}{\infty} \hat{F}(u) e^{-u \WeylLaplacian} \one
	\biggr]
	=
	\frac{1}{4\pi}
	\Tr
	\LimitInt{u}{0}{\infty} \hat{F}(u)
	\frac{1}{u}
	\biggl(
		1 + \frac{u}{6} \Ricci + \ldots
	\biggr)
,	
\label{eq:trace_anomaly_prelim}
\ee
where the ellipsis denotes terms higher order in the curvature and/or $\WeylLaplacian$. Recalling the condition~\eq{Universal-F}, it is immediately apparent that the expected result is recovered:
\be
	\Tr \anomaly = 
	\frac{1}{24 \pi}
	\Tr \Ricci
.
\ee

For completeness, let us check that the first term on the \rhs\ of~\eq{trace_anomaly_prelim} is convergent. Recalling~\eq{regularized_vacuum_flat} we may write:
\be
	\LimitInt{u}{0}{\infty} \hat{F}(u) e^{-u z}
	=
	z \der{}{z}
	\ln \frac{1 - \cutoff(z)}{1 - b\cutoff(z)}
.
\ee
Let us now integrate over $z$, from $v$ to $\infty$:
\be
	\LimitInt{u}{0}{\infty} \frac{\hat{F}(u)}{u} e^{-u v}
	=
	\int_{v}^\infty dz
	\,z \der{}{z}
	\ln \frac{1 - \cutoff(z)}{1 - b\cutoff(z)}
.
\ee
The integral on the \rhs\ has already been encountered below~\eq{regularized_vacuum_flat} and is known be to UV finite so the limit $v \rightarrow 0$ may be safely taken. Therefore,
\be
	\biggl\vert \LimitInt{u}{0}{\infty} \frac{\hat{F}(u)}{u} \biggr\vert < \infty
.
\ee

\subsection{The Polyakov Action}
\label{sec:Polyakov}

To extract the Polyakov action, which is the functional integral of the anomaly, we must integrate~\eq{VAC} in $\D=2$ to obtain the full Gaussian solution. Recalling~\eq{regularized_vacuum_flat} and the recipe described under~\eq{zd/dz}, we conclude that, formally
\begin{multline}
	\Stot^{\mathrm{Gauss}}_b[\dfield, \field, \chi, \overline{\chi}, g_{\mu\nu}]
	=
	\hf
	\dfield \cdot \ep^{-1}_b \cdot \dfield
	+
	\hf
	\field \cdot \ep^{-1}_1 \cdot \field
	+
	\overline{\chi} \cdot \ep^{-1}_1 \cdot \chi
\\
	-
	\Tr \ln \, \bigl[ (\one - \cutoff) \cdot (\one - b\cutoff)^{-1} \bigr]
	-\frac{1}{96\pi}
	 \Ricci \cdot \ep_0 \cdot \Ricci
	 +
	 C
,
\label{eq:FullGaussian}
\end{multline}
The
where $C$ is a potentially divergent topological invariant, to be discussed momentarily. penultimate term is the Polyakov action~\cite{PolyakovAction}; a derivation of the overall factor in which the sign conventions are the same as this paper can be found in~\cite{Codello-PolyakovAction}. While this term is non-local, it cancels a non-locality in the preceding term. The presence of the cutoff function ensures that the $\Tr \ln$ term is ultraviolet regularized. However, Taylor expanding the cutoff function it is apparent that there is a non-locality of the form $\Tr \ln (-\nabla^2)$.
That this contains a contribution which cancels the Polyakov action follows simply from the fact that
the anomaly cancels out in the differentiated expression~\eq{VAC}. Therefore, the Polyakov action must cancel out in the integrated expression~\eq{FullGaussian}.

As carefully discussed in~\cite{YellowPages}, $\Tr \ln (-\nabla^2)$ may contain a divergence on account of zero modes of the Laplacian. For example, on compact spaces without a boundary---for which the spectrum of the Laplacian is discrete---this is the case. The quantity, $C$, is chosen to remove a divergence of this type, should it exist. A more careful treatment, beyond the scope of this paper, would presumably mimic that of~\cite{YellowPages}, in which the field is `compactified' at intermediate stages of the calculation.

Returning to the Polyakov action, there is a particular limit of importance in which its cancellation does not hold. Transferring to dimensionful variables, instances of $-\nabla^2$ will become $-\nabla^2 / \Lambda^2$. For all $\Lambda > 0$, 
the non-local Polyakov action is cancelled, as it must be. However, for $\Lambda = 0$, the cutoff function vanishes (at least for non-zero modes), leaving just the non-local Polyakov action as the only vacuum contribution to the low energy effective action. Note that in cases where the spectrum of the Laplacian is discrete, and so the contribution of any zero modes to the trace is not of measure zero, the cancellation of the zero mode contributions by $C$ is necessary to ensure that no contributions from the $\Tr \ln $ term survive the limit $\Lambda \rightarrow 0$. However, it should be noted that additional, finite contributions to $C$ are not constrained within our approach.

This emergence of non-locality in the low energy limit is exactly analogous to the way correlation functions emerge in the presence of a source. For the Gaussian fixed-point (with $b=0$), the source-dependent action acquires a term which, in flat space, takes the form~\cite{Fundamentals}
\[
	\MomInt{p} J(p) J(-p) \frac{1-\cutoff(p^2/\Lambda^2)}{p^2}
.
\]
For all $\Lambda >0$, the action remains quasi-local. However, in the limit $\Lambda \rightarrow 0$ the cutoff function vanishes, leaving just physical $1/p^2$ correlation function.

\section{Legendre Transform}
\label{sec:Legendre}

\subsection{The Flow Equation}

For the particular choice of ERG equation used in this paper, the effective average action, $\Gint[\efield]$, is defined according to~\cite{HO-Remarks}
\be
	\Gint[\efield] = \Stot[\dfield] 
	-
	\hf
	\bigl(
		\dfield - \efield \cdot \cutoff
	\bigr)
	\cdot
	\cutoff^{-1} \cdot \Rinv
	\cdot
	\bigl(
		\dfield - \cutoff\cdot \efield
	\bigr)
,
\label{eq:EAA}
\ee
with $\Rinv$ defined by \eqs{Rinv}{R} and
$\efield$ determined by the condition:
\be
	\fderhat{\Stot[\dfield]}{\dfield}
	=
	\Rinv
	\cdot
	\bigl(
		\cutoff^{-1} \cdot \dfield  - \efield
	\bigr)
.
\label{eq:dS/dphi}
\ee
Given the flat-space form of the Legendre transform found in~\cite{CFPE},  the curved space generalization may be immediately deduced. However, it shall be explicitly derived in \app{Derivation}.
Either way, temporarily ignoring the regulator fields, and not explicitly  point-splitting, leads to
\be
	\biggl(	
		\scaling{\efield} \, \efield \cdot \fderhat{}{\efield}
		- 2 g_{\mu\nu} \cdot \fderhat{}{g_{\mu\nu}}
	\biggl)
	\Gint[\efield, g_{\alpha \beta}]
	=
	\frac{1}{2}
	 \Tr
	\biggl[
		G
		\cdot
		\Rinv
		\cdot
		\biggl(
			\bigl(
				\Gint^{(2)} + \Rinv \cdot \cutoff
			\bigr)^{-1}
			\cdot \Rinv
			-
			\cutoff^{-1}
		\biggr)
	\biggr]
	+ \Tr \anomaly
\label{eq:Legendre_pre}
\ee
with $\scaling{\efield} \equiv \scaling{\dfield}$ and
$\Gint^{(2)}(x,y) = \hat{\delta}^2 \Gint / \delta \Phi(x) \delta \Phi(y)$.
The functional trace defined such that, for some $F(x,y)$,  $\Tr F = \Int{x} \sqrt{g} F(x, x)$. 
A nice feature of working with the effective average action is that it is immediately apparent that taking an equal number of commuting and anti-commuting degrees of freedom, with the same value of $\eta$, causes cancellation of the $\Tr \cutoff^{-1}$ term. Better still, consider the effective average action for the free non-critical theories. Utilizing~\eq{NonCritical_TP} together with~\eq
{dS/dphi}, it immediately follows that, in these cases, $\efield = 0$. Therefore, if we suppose that each regulator field is indeed at the non-critical fixed-point with the appropriate value of $\eta$, the equation for the effective average action is simply
\be
	\biggl(	
		\scaling{\efield} \, \efield \cdot \fderhat{}{\efield}
		- 2 g_{\mu\nu} \cdot\fderhat{}{g_{\mu\nu}}
	\biggl)
	\Gint[\efield, g_{\alpha \beta}]
	=
	\frac{1}{2}
	\Tr
	\biggl[
		\Rinv \cdot \G \cdot \Rinv
		\cdot
		\bigl(
			\Gint^{(2)} + \Rinv \cdot \cutoff
		\bigr)^{-1}
	\biggr]
	+\Tr \anomaly
.
\label{eq:Legendre}
\ee
This, then, is the natural curved space generalization of the famous equation developed in~\cite{Bonini-1PI,Wetterich-1PI,Ellwanger-1PI,TRM-ApproxSolns}, specialized to a fixed-point. As with its progenitor, the structure of the functional trace ensures finiteness: the presence of $G$ provides UV regularization while $\Rinv$ acts like a mass (\cf~\eq{Rinv}), giving good IR behaviour. This justifies the comment made earlier that, in the Wilsonian effective action formulation, the regulator fields may always be considered to sit at the appropriate non-critical fixed-point.

It is straightforward to recover the Gaussian anomaly. First, though, we may justify the earlier statement that we may derive the fact that $\eta = 0$ for the Gaussian fixed-point. Let us restrict to quasi-local solutions of~\eq{Legendre} that are at most quadratic in $\efield$. It is apparent that a solution exists with $\Gamma^{(2)} = a \ep^{-1}_0$, together with $\eta = 0$. This defines the Gaussian solution, with positivity requiring that $a > 0$. For the Gaussian solution 
$\sigma = \ep^{-1}_0 \cdot (\one - \cutoff)^{-1}$, and so we recover the vacuum term on the \rhs\ of~\eq{GFP-GeneratedVac} (having taken the limit $y \rightarrow x$), and hence the anomaly, so long as we identify $a = 1/(1-b)$.

Following~\cite{HO-Remarks,CFPE}, we may write~\eq{Legendre} in a slightly different way by defining
\be
	\RIR \equiv \Rinv \cdot \cutoff
.
\ee
Recalling~\eq{intuition} and the discussion which follows, we define
\be
	\dot{\RIR} \equiv 
	-(\D+2)  \RIR - \delta_\weyl \RIR\bigl\vert_{\weyl = 1}
,
\ee
where the origin of the $\D$ can be traced to~\eq{Full-delta_weyl}.
Combining with~\eq{dR/dp^2}, implies that
\be
	\dot{\RIR} + \eta \RIR = \Rinv \cdot \G \cdot \Rinv
.
\ee
It is thus apparent that we may recast~\eq{Legendre} as
\be
	\biggl(	
		\scaling{\efield} \, \efield \cdot \fderhat{}{\efield}
		- 2 g_{\mu\nu} \cdot \fderhat{}{g_{\mu\nu}}
	\biggl)
	\Gint[\efield, g_{\alpha \beta}]
	=
	\frac{1}{2}
	\Tr
	\biggl[
		\bigl(\dot{\RIR} + \eta \RIR\bigr)
		\cdot	
		\bigl(
			\Gint^{(2)} + \RIR
		\bigr)^{-1}
	\biggr]
	+
	\Tr \anomaly
.
\label{eq:Legendre'}
\ee
In this context, it is natural to consider $\RIR$ to be an independent infra-red regulator, with dependence on $\eta$ now explicit~\cite{HO-Remarks}. Indeed, this makes it irresistible to propose a variant of~\eq{Legendre'} valid away from fixed-points:
\be
	\biggl(
		\partial_t + 
		\scaling{t} \, \efield \cdot \fderhat{}{\efield}
		- 2 g_{\mu\nu} \cdot \fderhat{}{g_{\mu\nu}}
	\biggl)
	\Gint_t[\efield, g_{\alpha \beta}]
	=
	\frac{1}{2}
	\Tr
	\biggl[
		\bigl(\dot{\RIR} + \eta(t) \RIR\bigr)
		\cdot
		\bigl(
			\Gint^{(2)}_t + \RIR
		\bigr)^{-1}
	\biggr]
	+\Tr \anomaly_t
,
\label{eq:Legendre_t}
\ee
where both the anomaly and the scaling dimension (equivalently $\eta$), may depend on $t$---which they must in the case of flows between two critical fixed-points with differing values of the scaling dimension and/or the anomaly.

\subsection{Beyond Global Weyl Transformations}
\label{sec:beyond}

Let us now address the question as to how, in flat space limit, conformal symmetry may arise. To this end, consider a generalization of~\eq{Legendre}:
\begin{multline}
	\weyl
	\cdot
	\biggl(	
		\scaling{\efield} \, \efield \times \fderhat{}{\efield}
		- 2 g_{\mu\nu} \times \fderhat{}{g_{\mu\nu}}
	\biggl)
	\Gint[\efield, g_{\alpha \beta}]
\\
	=
	\weyl \cdot
	\frac{1}{4}
	\Tr
	\biggl[
		\kernel{\Rinv}{\Rinv} \cdot
		\bigl(
			\Gint^{(2)} + \Rinv \cdot \cutoff
		\bigr)^{-1}
	\biggr]
\\
	+
	\CurvedInt{x}
	\Bigl(
		\weyl(x) \anomaly(x)
		+
		\partial_\mu \weyl(x)
		\mathcal{F}^\mu(\efield, g_{\alpha\beta})(x)
	\Bigr)
,
\label{eq:beyond}
\end{multline}
with $\Tr F(y,z; x) = \Int{y} \sqrt{g} F(y,y; x)$. For constant $\weyl$ it is clear that this reduces to~\eq{Legendre'}. In the flat space limit, the first terms on the \lhs\ and \rhs\ have been chosen to match 
the Legendre transformed version of~\eq{sct_condition}, as presented in~\cite{CFPE}. While the
anomaly will vanish in this limit, the final term will only vanish if $\mathcal{F}^\mu$ is a total derivative. This is guaranteed if there are no primary vector fields of dimension $\D-1$\footnote{Things are slightly more subtle in $\D=2$ as discussed in~\cite{Representations}}, recovering the usual condition for scale invariance to enhance to conformal invariance~\cite{Pol-ScaleConformal}.

In curved space, \eq{beyond} suggest how to go at least somewhat beyond global Weyl transformations. Supposing again that $\mathcal{F}^\mu$ is a total derivative, Weyl invariance is directly realized, up to the anomaly, as a property of the Wilsonian effective action for those $\weyl$ satisfying $\nabla^2 \weyl = 0$.

\subsection{The Trace of the Energy-Momentum Tensor}

Using the techniques of \app{Derivation} it is straightforward to derive the Legendre transform, $\emtLegendre$, of the trace of the energy-momentum tensor~\eq{trace_EMT}:
\be
	\emtLegendre
	=
	-\scaling{\efield}
	\biggl(
		\efield \times \fderhat{\Gamma}{\efield}
		-
		\RIR \cdot
		\bigl(
			\Gint^{(2)} + \RIR
		\bigr)^{-1}
	\biggr)
.
\ee
Compared to the Wilsonian effective action version, there is no need to perform point-splitting and, indeed, there is no analogue of the term where $\dfield$ is annihilated by a functional derivative. This serves to justify the claim that the cancellation of potentially divergent terms seen in \sect{Example} holds more generally.

\section{Conclusions}

A generalization of the ERG equation to curved space has been proposed, which explicitly includes the conformal anomaly and which reduces to the expected form in the flat space limit.
To arrive at this picture, non-critical fields must be included in order to regularize a vacuum term for which the ERG's in-built cutoff function does not help. These regulator fields are free but have vanishing two-point correlation functions: they are non-critical. 
Perhaps surprisingly, these fields essentially disappear from the effective average action. 
However, without the regulator fields, the Legendre transform from the Wilsonian effective action to the effective average action generates a divergent vacuum term. Carefully derived, the Legendre transformed equation~\eq{Legendre}---or, if one prefers, \eq{Legendre'}---provides a particularly concise statement of global Weyl invariance on a curved manifold, within a cutoff-regularized approach to quantum field theory. The latter equation has an appealing generalization to scale-dependent theories, given by~\eq{Legendre_t}.

A fuller understanding of local Weyl transformations within this framework is a natural extension. A step in this direction is provided by~\eq{beyond} which, in the flat space limit, reduces to the conformal fixed-point equation presented in~\cite{CFPE}.

\begin{acknowledgments}
	I would like to thank Hugh Osborn for useful correspondence and Tim Morris for helpful feedback.
\end{acknowledgments}

\appendix
\section{Legendre Transform Details}
\label{app:Derivation}

In this appendix, it is shown how to go from~\eq{Weyl-ERG} to~\eq{Legendre}, using~\eq{EAA}. 
From the latter it follows that
\be
	\fderhat{\Gint[\efield]}{\efield}
	=
	\Rinv
	\cdot
	\bigl(
		\dfield - \cutoff\cdot \efield
	\bigr)
.
\label{eq:dGamma/dPhi}
\ee
This equation shall be used to replace instances of $\dfield$; combining it with~\eq{dS/dphi} gives a recipe for replacing instances of $\hat\delta \Stot / \delta \dfield$:
\be
	\fderhat{\Stot[\dfield]}{\dfield}
	=
	\cutoff^{-1} \cdot 
	\fderhat{\Gint[\efield]}{\efield}
.
\ee
To start, we process the following set of terms:
\begin{multline}
	\scaling{\dfield} \, \dfield \cdot \fderhat{\Stot}{\dfield}
	+ \dfield \cdot \ep^{-1} \cdot \G \cdot {\fderhat{\Stot}{\dfield}}
	- \frac{1}{2}\fderhat{\Stot}{\dfield} \cdot \G \cdot {\fderhat{\Stot}{\dfield}}
\\ 
	=
	\scaling{\efield} \efield \cdot \fderhat{\Gint}{\efield}		
	+  \fderhat{\Gint}{\efield} \cdot \R
	\cdot
	\biggl(
	 	\scaling{\efield} 
		+ 
		\ep^{-1} \cdot G
	 \biggr) 
	 \cdot
	 \cutoff^{-1} \cdot \fderhat{\Gint}{\efield}
\\
	-\frac{1}{2}
	\fderhat{\Gint}{\efield} \cdot \cutoff^{-1} \cdot 
	G
	\cdot \cutoff^{-1} \cdot\fderhat{\Gint}{\efield}
	+\efield \cdot \cutoff \cdot \ep^{-1} \cdot
	G
	\cdot
	\cutoff^{-1} \cdot \fderhat{\Gint}{\efield}
.
\label{eq:EAA-process}
\end{multline}
Many of these terms will cancel ultimately against those generated from varying the metric. With this in mind, we utilize a combination of the chain rule, \eqs{EAA}{dGamma/dPhi} to derive
\begin{multline}
	-2 \CurvedInt{x}  g_{\mu\nu}(x)
	\fderhat{\Stot[\dfield]}{g_{\mu\nu}(x)}\biggr\vert_{\dfield}
	=
	-2 \CurvedInt{x} g_{\mu\nu}(x)
	\biggl\{
\\
		\fderhat{\Gint[\efield]}{g_{\mu\nu}(x)}\biggr			
		\vert_{\efield}
		+
		\fderhat{\Gint}{\efield} \cdot \fderhat{\efield}{g_{\mu\nu}(x)} \biggr\vert_{\dfield}
		+
		\hf
		\fderhat{}{g_{\mu\nu}(x)} \biggr\vert_{\dfield}
		\biggl(
			\fderhat{\Gint}{\efield} \cdot \R \cdot \cutoff^{-1} \cdot \fderhat{\Gint}{\efield}
		\biggr)
	\biggr\}
.
\label{eq:dS/dg-process}
\end{multline}
The middle term on the \rhs\ can now be processed using~\eq{dGamma/dPhi}, whereupon one of the resulting contributions cancels the last term in~\eq{dS/dg-process}. To see this, it is convenient to first insert unity, in the form $\cutoff^{-1} \cdot \cutoff$. Recalling the effects of infinitesimal Weyl transformations, \eqs{weyl-g}{delta_weyl} it is apparent that
\begin{align}
	-2 \CurvedInt{x} \weyl(x) g_{\mu\nu}(x)
	\fderhat{\Gint}{\efield} \cdot \fderhat{\efield}{g_{\mu\nu}(x)} \biggr\vert_{\dfield}
	\hspace{-30ex}
\nonumber
\\
	 & =
	-2 \CurvedInt{x} \weyl(x) g_{\mu\nu}(x)
	\fderhat{\Gint}{\efield} \cdot \cutoff^{-1} \cdot
	\fderhat{\cutoff \cdot \efield}{g_{\mu\nu}(x)} \biggr\vert_{\dfield}
	+
	\fderhat{\Gint}{\efield} \cdot \cutoff^{-1} \cdot
	\delta_\weyl\bigl( \cutoff \cdot  \bigr) \efield
\nonumber
\\
	&
	=
	2 \CurvedInt{x} \weyl(x) g_{\mu\nu}(x)
	\CurvedInt{y}
	\biggl(
		\fderhat{\Gint}{\efield} 
		\cdot
		\cutoff^{-1}
	\biggr)(y)
	\fderhat{}{g_{\mu\nu}(x)}\biggr\vert_{\dfield}
	\biggl(
		\R \cdot \fderhat{\Gint}{\efield}
	\biggr)(y)
\nonumber
\\
	& 
	\qquad \qquad
	+ \fderhat{\Gint}{\efield} \cdot \cutoff^{-1} \cdot
	\delta_\weyl\bigl( \cutoff \cdot  \bigr) \efield
\nonumber
\\
	&	
	=
	\CurvedInt{x} \weyl(x) g_{\mu\nu}(x) \fderhat{}{g_{\mu\nu}(x)} \biggr\vert_{\dfield}
	\biggl(
		\fderhat{\Gint}{\efield} \cdot \cutoff^{-1} \cdot \R  \cdot \fderhat{\Gint}{\efield}
	\biggr)
	-\frac{\D}{2}
	\fderhat{\Gint}{\efield}\cdot \cutoff^{-1} \weyl  \cdot \R 
	\cdot \fderhat{\Gint}{\efield}
\nonumber
\\
	&
	\qquad
	+
	\hf
	\fderhat{\Gint}{\efield} \cdot 
	\Bigl(
		\cutoff^{-1} \cdot 
		\delta_\weyl \bigl(  \R \cdot   \bigr)
		-\delta_\weyl \bigl( \cutoff^{-1} \cdot  \bigr) \R \cdot 
	\Bigr)
	\fderhat{\Gint}{\efield}
	+
	\fderhat{\Gint}{\efield} \cdot \cutoff^{-1} \cdot
	\delta_\weyl\bigl( \cutoff \cdot  \bigr) \efield
.
\label{eq:dS/dg-detail}
\end{align}
A word about notation is required. Recall~\eqss{one}{dot_notation-Curved}{K(x,y)-Curved}. From this we see that
\be
	\bigl(\cutoff \cdot \efield\bigr)(x)
	=
	\cutoff(\WeylLaplacian) \efield(x)
\ee
and so we may understand $\delta_\weyl(\cutoff \cdot)$ to reduce to $\delta_\weyl \cutoff(\WeylLaplacian)$, with the inclusion of the dot within the scope of $\delta_\weyl$ emphasising that the factor of $1/\sqrt{g}$ coming from the $\delta$-function is cancelled by the factor of $\sqrt{g}$ coming from the integral. Returning to~\eq{dS/dg-detail} , since $\cutoff^{-1} \cdot  \R = \R \cdot \cutoff^{-1}$, the first term on the \rhs\ exactly removes the last term in~\eq{dS/dg-process}.

The next step is to collect together surviving terms of a common form from~\eqs{EAA-process}{dS/dg-process}, making use of~\eq{dS/dg-detail}. First, we examine terms $\order{\efield \,\delta \Gint /\delta \efield}$, the sum of which we shall denote by $\mathcal{C}[\efield]$:
\be
	\mathcal{C}[\efield]
	=
	\scaling{\efield} \efield \cdot \fderhat{\Gint}{\efield}
	+
	\efield
	\cdot
	\ep_0^{-1} \cdot G
	\cdot 
	\cutoff^{-1}
	\cdot
	\fderhat{\Gint}{\efield}
	+
	\fderhat{\Gint}{\efield} \cdot \cutoff^{-1} \cdot
	\delta_\weyl\bigl( \cutoff \cdot  \bigr) \efield \biggr\vert_{\weyl = 1}
.
\label{eq:C-terms}
\ee
To proceed, use~\eqs{intuition}{G(p^2)} to recast
\be
	\delta_\weyl \cutoff(\WeylLaplacian)\Bigr\vert_{\weyl = 1}
	 = 
	- \LimitInt{u}{0}{\infty} \hat{G}(u)\,
	e^{-u\WeylLaplacian} \WeylLaplacian 
,
\label{eq:delta_w-K}
\ee
from which it follows that
\be
	\fderhat{\Gint}{\efield} \cdot \cutoff^{-1} \cdot
	\delta_\weyl\bigl( \cutoff \cdot  \bigr) \efield
\\
	 = 
 	-\fderhat{\Gint}{\efield} \cdot \cutoff^{-1} \cdot
	G
	\cdot
 	\efield
\ee
The first term, when inserted into the second line of~\eq{C-terms} cancels the following term. Therefore,
\be
	\mathcal{C}[\efield]
	=
	\scaling{\efield}
	\efield \cdot	
	\fderhat{\Gint}{\efield}
.
\ee

Now we go back to~\eqss{EAA-process}{dS/dg-process}{dS/dg-detail} and analyse the terms of the schematic form $(\delta \Gint / \delta \efield)^2$, which shall be denoted by $\mathcal{D}\efield]$:
\begin{multline}
	\mathcal{D}[\efield] =
	\fderhat{\Gint}{\efield}
	\cdot
	\biggl[
		\R
		\cdot
		\bigl(
		 	\scaling{\efield}
			+ 
			\ep^{-1} \cdot G
		 \bigr) 
		 \cdot
		 \cutoff^{-1}
		-\frac{1}{2}
		 \cutoff^{-1} \cdot G
		\cdot \cutoff^{-1} 
		-\frac{\D}{2}
		\cutoff^{-1} \cdot \R
	\biggr]
	\cdot \fderhat{\Gint}{\efield}
\\
	+\hf
	\fderhat{\Gint}{\efield}
	\cdot
	\Bigl(
		\cutoff^{-1} \cdot 
		\delta_\weyl \bigl(  \R \cdot   \bigr)
		-\delta_\weyl \bigl( \cutoff^{-1} \cdot  \bigr) \R \cdot 
	\Bigr)
	\fderhat{\Gint}{\efield}
	\biggr\vert_{\weyl = 1}
.
\label{eq:TreeTerms}
\end{multline}
Observing that the first term on the \rhs\ of~\eq{TreeTerms} is partially cancelled by the $\D/2$ term yields
\begin{multline}
	\mathcal{D}[\efield] = 
	\hf
	\fderhat{\Gint}{\efield}
	\cdot
	\Bigl(
		\cutoff^{-1} \cdot 
		\delta_\weyl \bigl(  \R \cdot   \bigr)
		+ \cutoff^{-1} \cdot \delta_\weyl \bigl( \cutoff \cdot  \bigr) \cdot \cutoff^{-1} \cdot \R \cdot 
	\Bigr)
	\fderhat{\Gint}{\efield}
	\biggr\vert_{\weyl = 1}
\\
	+
	\hf
	\fderhat{\Gint}{\efield}
	\cdot
	\cutoff^{-1}
	\cdot
	\biggl\{
		(\eta - 2)
		\R 
		+
		2 G
		\cdot
		\ep^{-1} \cdot \R
		+
		G
		\cdot
		\cutoff^{-1}
	\biggr\}
	\cdot
	\fderhat{\Gint}{\efield}
.
\label{eq:D-terms}
\end{multline}
The second term may be processed by using~\eq{delta_w-K}. The treatment of the first term follows similarly.
\begin{align}
	\delta_w \R(\WeylLaplacian) \Bigl\vert_{\omega = 1}
	& =
	2\LimitInt{u}{0}{\infty} u \hat{\R}(u)
	e^{-u\WeylLaplacian} \WeylLaplacian 
.
\label{eq:delta_weyl_rho}
\end{align}
Utilizing~\eq{dR/dp^2} and recalling the discussion under~\eq{intuition} we deduce that
\be
	\delta_w \R
	\Bigl\vert_{\weyl =1}
	=
	(2- \eta ) \R
	-
	\G \cdot  \ep^{-1} \cdot \R
	+
	\G \cdot \cutoff^{-1}
\ee
and so it is apparent that  $\mathcal{D}[\efield]$ vanishes. Note that, if we so desired, we could define $\R$ such that this is true, providing the alternative justification of~\eq{R} promised earlier.

To complete the Legendre transform, all that remains is to treat
\be
	\mathcal{Q}[\efield]
	=
	\frac{1}{2}
	\fderhat{}{\dfield} \cdot \G \cdot \fderhat{\Stot}{\dfield}
.
\ee
To do so we differentiate~\eq{dS/dphi} \wrt\ $\dfield$ and~\eq{dGamma/dPhi} \wrt\ $\efield$:
\begin{subequations}
\begin{align}
	\biggl(\R \cdot \fderhat{}{\dfield}\biggr)(x)
	\;
	\fderhat{\Stot}{\dfield(y)} - \cutoff^{-1}(x,y) & = 	
	-\fderhat{\efield(x)}{\dfield(y)}
,
\\
	\biggl(\R \cdot \fderhat{}{\efield}\biggr)(x)
	\; 
	\fderhat{\Gint}{\efield(y)} + \cutoff(x,y) & = \fderhat{\dfield(x)}{\efield(y)}
.
\end{align}
\end{subequations}
From this it follows that
\be
	\dfderhat{\Stot}{\dfield}{\dfield}
	=
	-\Rinv
	\cdot
	\biggl(
		\dfderhat{\Gint}{\efield}{\efield} + \Rinv \cdot \cutoff
	\biggr)^{-1}
	\cdot \Rinv
	+
	\Rinv \cdot \cutoff^{-1}
.
\ee
Putting everything together yields~\eq{Legendre_pre}.


\end{document}